\documentstyle[aps,prl,epsfig]{revtex}
\begin{document}


\title{Non-Gaussian  equilibrium in  a long-range Hamiltonian system}

\author{Vito Latora$^a$, Andrea Rapisarda$^b$  and 
Constantino Tsallis$^c$ }

\address{$^a$Dipartimento di Fisica e Astronomia,  Universit\'a di Catania,
and INFN sezione di Catania,\\
Corso Italia 57 I 95129 Catania, Italy}

\address{$^b$  Laboratoire de Physique Th\`eéorique et Mod\`eèles Statistiques,
  Universit\'e
 Paris-Sud, 91405 Orsay Cedex, France }

\address{$^c$ Centro Brasileiro de Pesquisas Fisicas,  Rua Xavier Sigaud 150,
 22290-180
Rio de Janeiro, Brazil}

\date{\today}
\maketitle

\begin{abstract}
We study the dynamics of a system of N classical 
spins with infinite-range interaction. 
We show that, if  the thermodynamic limit is taken before the 
infinite-time limit, the system does not relax to the 
Boltzmann-Gibbs equilibrium, 
but exhibits  different equilibrium properties, 
characterized by stable non-Gaussian velocity distributions, 
L\'evy walks and dynamical correlation in phase-space.\\

\end{abstract}
 
 

\bigskip
Though not always clearly stated, standard equilibrium 
thermodynamics \cite{pat,stan,land} is valid
 only for sufficiently short-range interactions.  
This is not the case, for example,
for gravitational or unscreened Coulombian fields, 
or for systems with long-range microscopic memory 
and fractal structures in phase space.  
The increasing experimental
evidence of dynamics and thermodynamics anomalies 
in turbulent plasmas \cite{bog} and fluids \cite{bec,sol,shl} ,
astrophysical systems \cite{lyn,pie,pos,kon,tor}, 
nuclei \cite{gro,dago} and atomic clusters\cite{ato}, 
granular media\cite{kud},  glasses\cite{par,stil}
and complex systems\cite{idra,sta1} found in the last years, 
provide further motivation  for a generalization of thermodynamics.     

In this paper we consider a simple model of classical
spins with infinite range interactions\cite{ant1,lat1,lat2,lat3}, 
and we show that, if the thermodynamic limit is
performed before the infinite time limit, 
the system does not relax to the Boltzmann-Gibbs (BG)
equilibrium, but exhibits different equilibrium properties  
characterized by non-Gaussian velocity distributions, 
L\'evy walks and dynamical  correlation in phase-space, 
and the validity of the zeroth principle of thermodynamics. 
Our results show some consistency with the predictions of a generalized 
non-extensive thermodynamics recently proposed\cite{tsa1,tsa3}.
The Hamiltonian Mean Field  (HMF)  model describes a system of 
N planar classical spins interacting through an 
infinite-range potential\cite{ant1}.  
The Hamiltonian can be written as:
\begin{equation}
        H=K+V= \sum_{i=1}^N  {{p_i}^2 \over 2} +
  {1\over{2N}} \sum_{i,j=1}^N  [1-cos(\theta_i -\theta_j)]~~,
\end{equation}
\noindent
 where $\theta_i$ is the $ith$ angle and $p_i$ the 
conjugate variable   representing   the  angular momentum
(or the rotational velocity since unit mass is assumed). 
The interaction is the same as in the
ferromagnetic X-Y model \cite{stan}, though the 
summation is extended to all couples of spins and not
restricted to first neighbors. 
Following tradition, the coupling constant in the potential is 
divided by N. This makes H only formally extensive 
($V\sim N$ when $N\rightarrow\infty$)\cite{tsa1,tsa3,celia,gia}, 
since the energy remains non-additive, i.e. the system cannot 
be trivially divided in two independent sub-systems.   
The canonical analytical solution of the model 
predicts a second-order phase transition from a low-energy 
ferromagnetic phase with magnetization  $M\sim1$  
(M is the modulus of  ${\bf M}={\frac{1}{N}}\sum_{i=1}^N {\bf m}_i$ , 
where ${\bf m}_i=[cos(\theta_i), sin(\theta_i)]$, 
to a high-energy one where the spins are homogeneously oriented 
on   the unit circle and $M\sim0$. 
The {\em caloric curve}, i.e. the  
dependence of  the energy density $U = E/N$ on the temperature $T$,  
is  given by $U = {T \over 2} +  {1 \over 2} ( 1 - M^2 )$ 
and shown in Fig.1(a). 
The critical point is  at energy density $U_c=0.75$ 
corresponding to a critical temperature $T_c=0.5$ \cite{ant1}. 
The dynamical behavior of HMF can be investigated in the microcanonical 
ensemble by starting the system with water bag initial conditions (WBIC), 
i.e. $\theta_i=0$  for all $i$ ($M=1$) and velocities uniformly 
distributed, and integrating numerically the equations of motion \cite{lat1}. 
As shown in Fig.1(a), microcanonical simulations are in general in good
agreement with the canonical ensemble, except for a region below $U_c$, 
where  it has also been  found a
dynamics characterized by  L\'evy walks, anomalous diffusion \cite{lat2}  
and a negative specific heat\cite{lat3}.
Ensemble inequivalence and negative specific heat 
have also been found in self-gravitating systems \cite{lyn}, 
nuclei and atomic clusters  \cite{gro,dago,ato},
though in the present model such anomalies emerge as dynamical 
features  \cite{ant2,lat4}. 
In order to understand better  this disagreement we focus on a particular 
energy value, namely $ U=0.69$, and we follow the time evolution 
of temperature, magnetization, and velocity distributions.
              
In Fig.1(b) we report the time evolution of $2<K>/N$, 
a quantity that, evaluated at equilibrium, 
is expected to coincide with the temperature 
($<\cdot>$ denotes time averages). The system is started 
with WBIC and rapidly reaches a metastable or quasi-stationary 
state (QSS) which does not coincide with   the canonical prediction. 
In fact, after a short transient time, $2<K>/N$ shows 
a plateau corresponding to a N-dependent temperature 
$T_{QSS}(N)$ (and $M_{QSS}\sim 0$) lower than the canonical temperature.  
This metastable state needs a long time to relax to the 
canonical equilibrium state with temperature 
$T_{can}=0.476$ and magnetization $M_{can}=0.307$. 
The duration of the plateau increases with the size of the system: in
particular we have checked that the lifetime of QSS has 
a linear dependence on N, see Fig.1(c). 
Therefore the two limits  $ t\rightarrow \infty$ and $ N\rightarrow \infty$ 
do not commute and if  the thermodynamic limit 
is performed before the infinite time limit, 
the system does not relax to the BG equilibrium. 
This has been conjectured to be an ubiquitous feature 
in non-extensive systems\cite{tsa1}, 
but it has also been found for spin glasses\cite{par}.  
When N increases $T_{QSS}(N)$ tends to $T_{\infty} =0.380$, a value
obtained analytically as the metastable prolongation 
(at energies below $U_c=0.75$) of the
 high-energy solution ($M=0$). 
We have also found that $ [T_{QSS}(N)-T_{\infty}] \propto N^{-1/3}$ 
and $M_{QSS} \propto  N^{-1/6}$, see Fig. 1(d). 
At the  same time we have checked that increasing the
size, the largest Lyapunov exponent for the QSS tends to zero.  
In this sense mixing is negligible and one expects anomalies 
in the relaxation process \cite{kry}.
The fact that $T_{QSS}$  converges to a nonzero value of 
temperature for $N\rightarrow \infty$ 
means that, when N is macroscopically large, 
systems can share the same temperature, though this equilibrium 
is not the familiar one.  
All this amounts to say that the zeroth principle 
of thermodynamics is stronger than what one might think 
through BG statistical mechanics, 
since it is true even when the system is 
not at the usual BG equilibrium. We have checked the robustness 
of the above results  by changing the level of 
accuracy of the numerical integration and 
by adding  small perturbations. 
We also verified that the QSS has a finite basin of attraction, 
by adopting different  initial conditions,  as for example  double 
water bag  (DWBIC). 
In Fig.2 we focus on the velocity probability distribution 
functions (pdfs). The initial velocity pdfs (WBIC or DWBIC) , 
reported in Fig. 2(a) , quickly acquire and maintain 
during the entire duration of the metastable state 
a  {\em non-Gaussian  shape }, see Figs.2(b) and 2(c).  
The velocity pdf of the QSS is wider than a Gaussian 
for small velocities, but shows a faster decrease for $p>1.2$.
The enhancement for velocities around $ p\sim 1$  is   
consistent with the anomalous diffusion and 
the L\'evy walks (with average velocity $ p\sim 1$) 
 observed in the QSS regime \cite{lat2}. 
The following rapid decrease for $p>1.2$ is due to conservation of 
total energy . 
The stability of the QSS velocity pdf can be explained by the fact 
that, for $N\rightarrow \infty$, 
  $M_{QSS} \rightarrow 0$  and thus  the force on  
the spins tends to zero with N, being $F_i = -M_x sin\theta_i + M_y cos\theta_i$. 
Of course,  for finite N,  we have always 
 a small random force, which makes the system  eventually 
evolve into the usual Maxwell-Boltzmann
 distribution after some time. We show this for small systems (N=500,1000) 
at time  t=500000 in Fig.2(e).  When this happens, 
  L\'evy walks disappear and anomalous 
diffusion leaves place to  Brownian diffusion\cite{lat2} .
A possible frame to reproduce the non-Gaussian pdf in Fig.2 (b) could be 
the non-extensive statistical mechanics recently proposed\cite{tsa1,tsa3}
with the entropic index $q \ne 1$. 
This formalism provides, for the canonical ensemble,  a 
q-dependent power-law distribution in the variables $p_i$ , $ \theta_i$ . This 
distribution  has  to be integrated over all $\theta_i$ and all but
 one $p_i$ in order to obtain the one-momentum pdf,
  $P_q(p)$, to be compared with the  numerical  one, $P_{num}(p)$, 
obtained by considering, 
within the present molecular dynamical frame, increasingly large N-sized subsystems 
of an increasingly large M-system. Within the $M>> N >>1$
 numerical limit, we expect to go from 
the microcanonical ensemble to the canonical one 
(the cut-off is then expected to gradually disappear as indeed 
occurs in the usual short-range Hamiltonians), thus 
justifying the comparison between $P_q(p)$ and $P_{num}(p)$. 
The enormous complexity of this procedure
 made us to turn instead onto a naive, but tractable, comparison,
 namely that of our present
 numerical results with the following one-free-particle pdf \cite{tsa1}    
 $ P(p) = [1 - ({1 \over 2T}) (1-q) p^2 ]^{ 1/(1-q)}$  , which
 recovers the Maxwell-Boltzmann distribution for $q=1$. This formula 
 has been recently used to describe successfully turbulent 
 Couette-Taylor flow \cite{bec}
 and non-Gaussian pdfs related to anomalous diffusion of  {\em Hydra} cells in
 cellular aggregates \cite{idra}. In our case,
 the best fit is obtained by a curve with $q=7$, $T=0.38$ 
as shown in Figs. 2 (b) and 2 (c). 
The agreement between numerical results and theoretical curve improves 
with the size of the system. A finite-size scaling confirming  
the validity of the fit is reported in panel (d), where $\Delta=P_ {th} - P_{num} $, 
the difference between the numerical results and the theoretical curve for $q=7$,  
is shown to go to zero as a power of N (for four values of p). 
Since $q>3$,  the theoretical curve does 
not have a finite integral and therefore it needs to be truncated 
with a sharp cut-off to make the total probability equal to one.
It is however clear that, the fitting 
value $q=7$ is only an effective non-extensive entropic index. 
Similar non-Gaussian pdfs have also been found in 
turbulence and granular matter  experiments 
\cite{bec,kud}, though this is the first evidence 
in a  Hamiltonian system.
In Fig.3 we verify, through the calculation of the fractal  
dimension $D_2$ \cite{gras},  that  a 
dynamical correlation emerges in the $\mu$-space  before the final
 arrival to a quasi-uniform 
distribution. During intermediate times some filamentary structures 
appear, a similar feature
 has recently been found also in self-gravitating systems\cite{kon}, 
which might be closely related to
 the plateaux observed in Fig.1(b). We learn from the curves in Fig.3(c) that,
 since they do not 
sensibly depend on N, the possible connection does  not concern the entire 
$\mu$-space, but perhaps
 only the small  sticky regions between the "chaotic sea" and 
the quasi-orbits\cite{kla}.

 Metastable states are 
ubiquitous in nature. Their full understanding is, however, far from
 trivial. They basically correspond 
to local, instead of global, minima of the relevant 
thermodynamic energy. The two types of minima are separated 
by activation barriers which, 
at the thermodynamic limit, can be low, high or infinite, 
all of them presumably occurring
 in nature. The last case yields of course to quite drastic
 consequences. Moreover, the local
 minimum can either  make 
the system  to live in a smooth part of the a priori accessible
 phase space, or it can force it to live in a geometrically more complex 
(e.g., multifractal)
 part of the phase space. The richness of 
such situation is what makes interesting
 the study of
 glasses, nuclei, atomic clusters, 
self-gravitating  and other complex systems. It is natural
 to expect for such systems 
that the infinite size and infinite time limits are not 
interchangeable. What has emerged quite clearly 
here is that thermodynamically large systems 
with long-range interactions belong to this 
very rich class. We have verified that the usual 
attributes of thermal equilibrium: 
{\em zeroth principle at finite temperatures, robustness associated 
with a finite basin of 
attraction in the space of the initial conditions, stable distribution of
 velocities},  are satisfied, 
but they  {\em systematically differ from what BG statistical mechanics
 has make familiar to us 
along the last 130 years}. 
Our findings indicate some consistency with 
the predictions of non-extensive statistical mechanics \cite{tsa1}, 
though a firm and unambiguous connection remains a challenge 
for future studies. 
In particular we  believe all these features not to be
 exclusive of the present HMF model. 
Similar scenarios are expected for systems with say
 two-body interactions decaying   like  $r^{-\alpha}$   for   
$0 \le \alpha \le \alpha_c$ , where  $\alpha_c$  is equal, for 
classical systems, to the space dimension \cite{celia,gia}.

\vskip 0.5truecm
\noindent
We thank M. Antoni,
 F. Baldovin, M. Baranger, E.P. Borges, E.G.D. Cohen, X. Campi, H. Krivine, 
M. M\'ezard, S. Ruffo and A. Torcini for stimulating discussions.

\bigskip
\noindent
E-mail: vito.latora@ct.infn.it

\noindent
E-mail: andrea.rapisarda@ct.infn.it

\noindent
E-mail: tsallis@cbpf.br


\begin{figure}
\begin{center}
\epsfig{figure=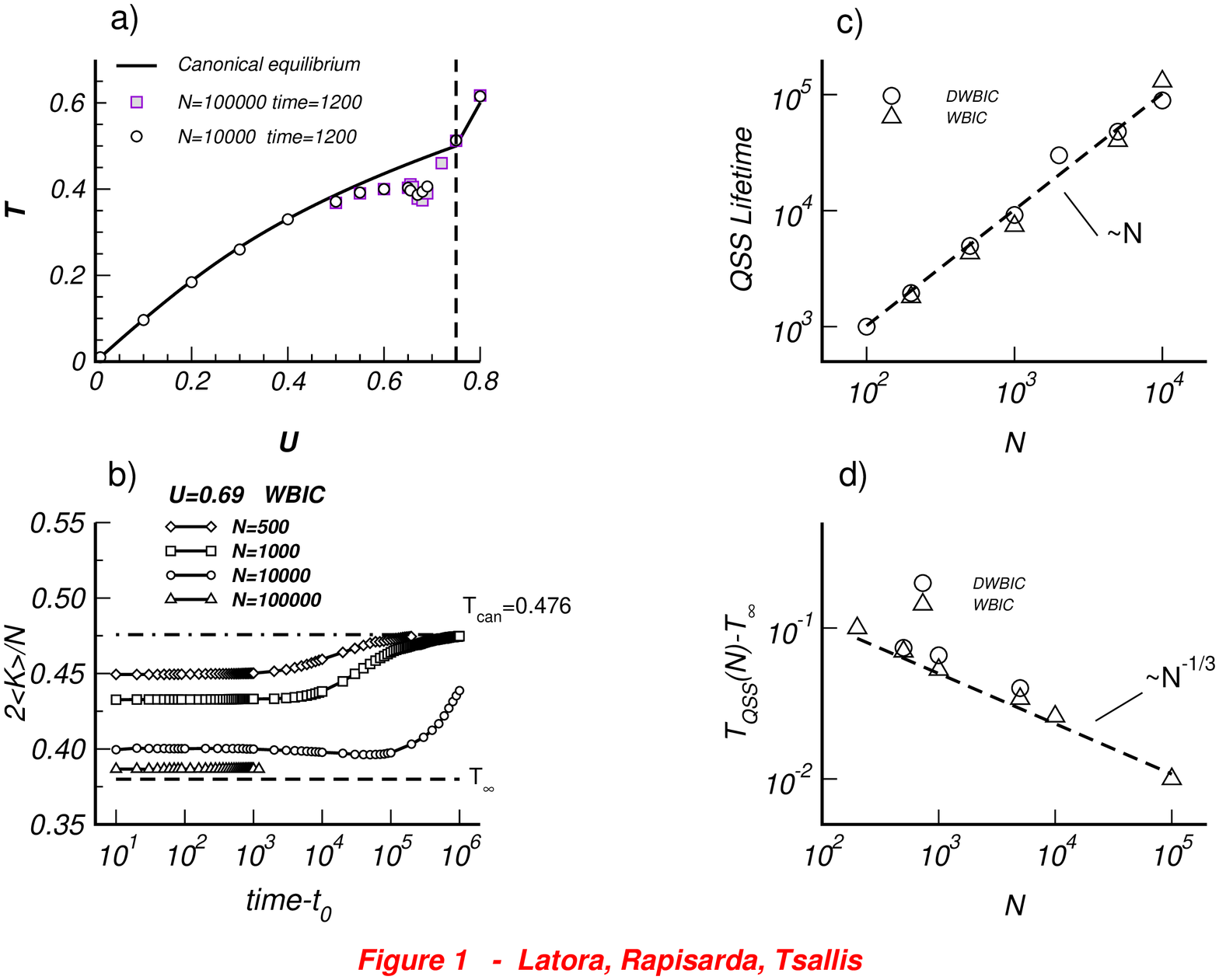,width=12truecm,angle=-90}
\end{center}
\caption{ 
(a) Caloric curve: microcanonical numerical results for $N=10000, 100000$ 
are compared with equilibrium theory in the  canonical ensemble. 
The dashed vertical line indicates the critical energy. Water bag
initial conditions   
(WBIC) are used in the numerical simulations.  
Temperature is computed from $T=2<K>/N$, 
where $<\cdot·>$ denotes time averages after a short transient 
time $t_0=10^2$ (not reported here).  
(b) Microcanonical time evolution of $2<K>/N$, for the energy density 
 $U=0.69$ and different sizes.  
Each curve is an average over 
typically $100-1000$ events. The dot-dashed line represents 
the canonical temperature $T_{can}=0.476$. The quantity $2<K>/N$
which starts from an initial value $1.38$ ($V=0$ and $K=U\cdot N$ in WBIC), 
does not relax immediately to the canonical temperature. 
The system lives in a quasi-stationary state (QSS)
with a plateau {\em temperature} 
$T_{QSS}(N)$ smaller than the expected value $0.476$.   Lifetime of QSS
increases with N and the 
value of their temperature converges, as $N$ increases, to  the 
temperature $ T_{\infty}=0.38$, 
reported as a  dashed line. 
Log-log plot of QSS lifetime (c) and 
$T_{QSS}(N)-T_{\infty}$ (d) are reported 
as a function of the size $N$. 
 Lifetime diverges linearly with $N$, and  $T_{QSS}(N)$ converges to 
$T_{\infty}=0.38$ as $N^{-1/3}$ (see fit shown as a dashed line). 
Note that from the caloric curve one gets
$ M^2=T+1-2U=T-0.38$. 
Therefore from the behaviour reported in panel (d), being $ T_{\infty}=0.38$, 
one gets $M_{QSS}=  N^{-1/6}$. Results are similar 
when we consider double water bag initial conditions (DWBIC),
i.e. $\theta_i=0$  for all $i$ 
and velocities uniformly distributed in ($-p_2,-p_1$) and ($p_1,p_2$).
In figure we report the case $p_1=0.8$, $p_2=1.51$. }
\end{figure}

\begin{figure}
\begin{center}
\epsfig{figure=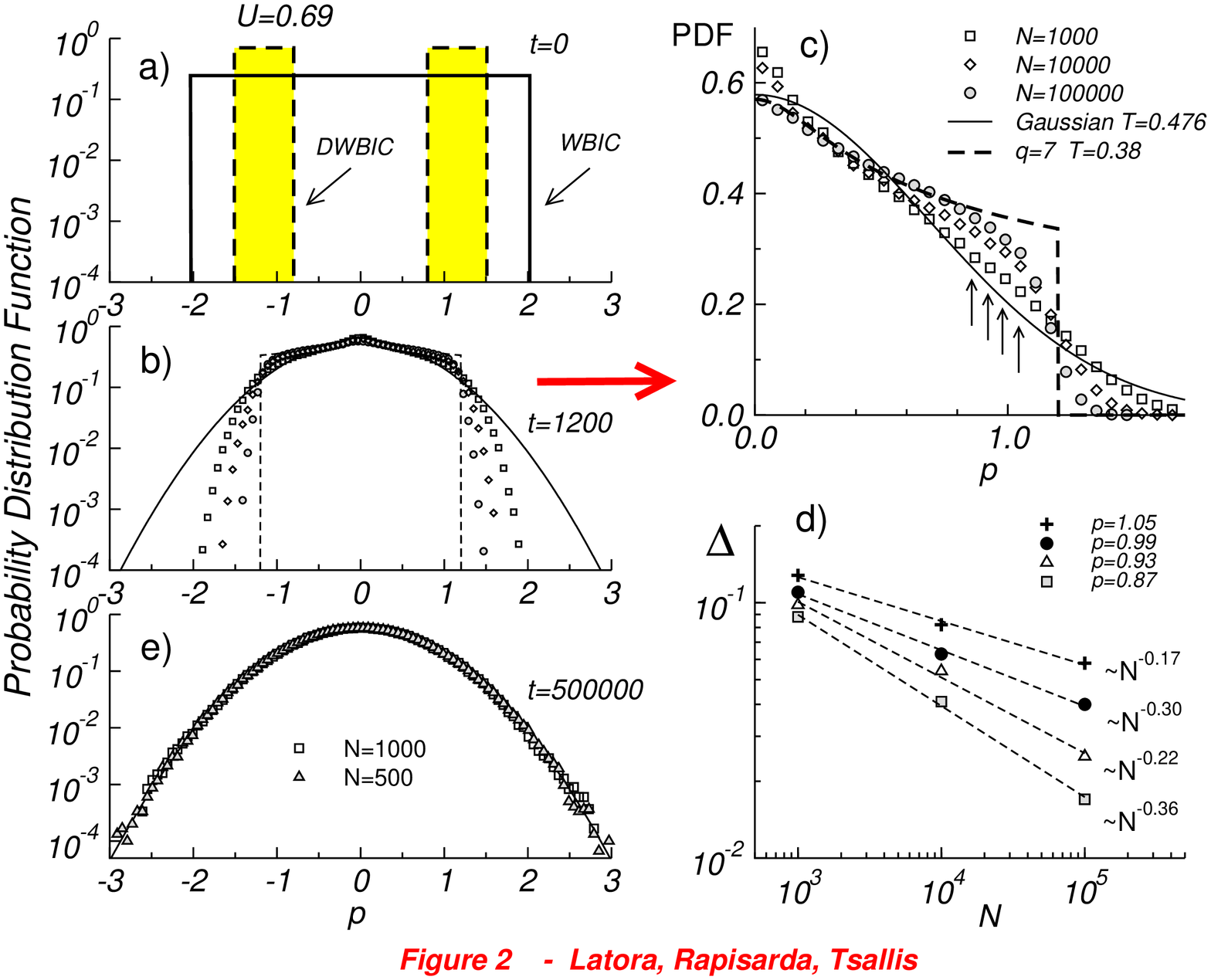,width=11truecm,angle=-90}
\end{center}
\caption{ Time evolution of the velocity probability 
distribution function (pdf)  for $ U=0.69$ and 
different sizes of the system.  
(a) At time $t=0$ we start with a single (WBIC) or a double (DWBIC) 
water bag velocity pdf. 
(b) In the transient regime where $2<K>/N$ shows a  plateau corresponding to 
$T_{QSS}(N)$ and  the system 
lives in a Quasi-Stationary State (QSS), the velocity pdfs 
do not change in time and are very different from the Gaussian canonical equilibrium 
distribution (full curve). 
The pdfs at time  $t=1200$ for $N=1000,10000,100000$ show a 
convergence towards a non-Gaussian  distribution which can be fitted by
means of a  power-law analytical curve (dashed curve) 
consistent with the generalized non-extensive thermodynamics \protect\cite{tsa1}
proposed by Tsallis and characterized by $q=7$ and $T=0.38$, see text. 
The theoretical curve has been truncated with a sharp cut-off in order  to 
have total probability equal to one, see text. 
(c) The same curves shown in (b) at $t=1200$ are reported in linear scale. 
(d) We show the difference, $\Delta$, between the theoretical curve
and the numerical results, 
as function of N for the four values of p indicated by arrows in panel (c). 
(e) We show  the numerical pfds at $t=500000$ for  $N=500$ and $1000$. We get 
an excellent  agreement with the Gaussian  canonical equilibrium distribution 
at temperature $T=0.476$.   
}
\end{figure}

\begin{figure}
\begin{center}
\epsfig{figure=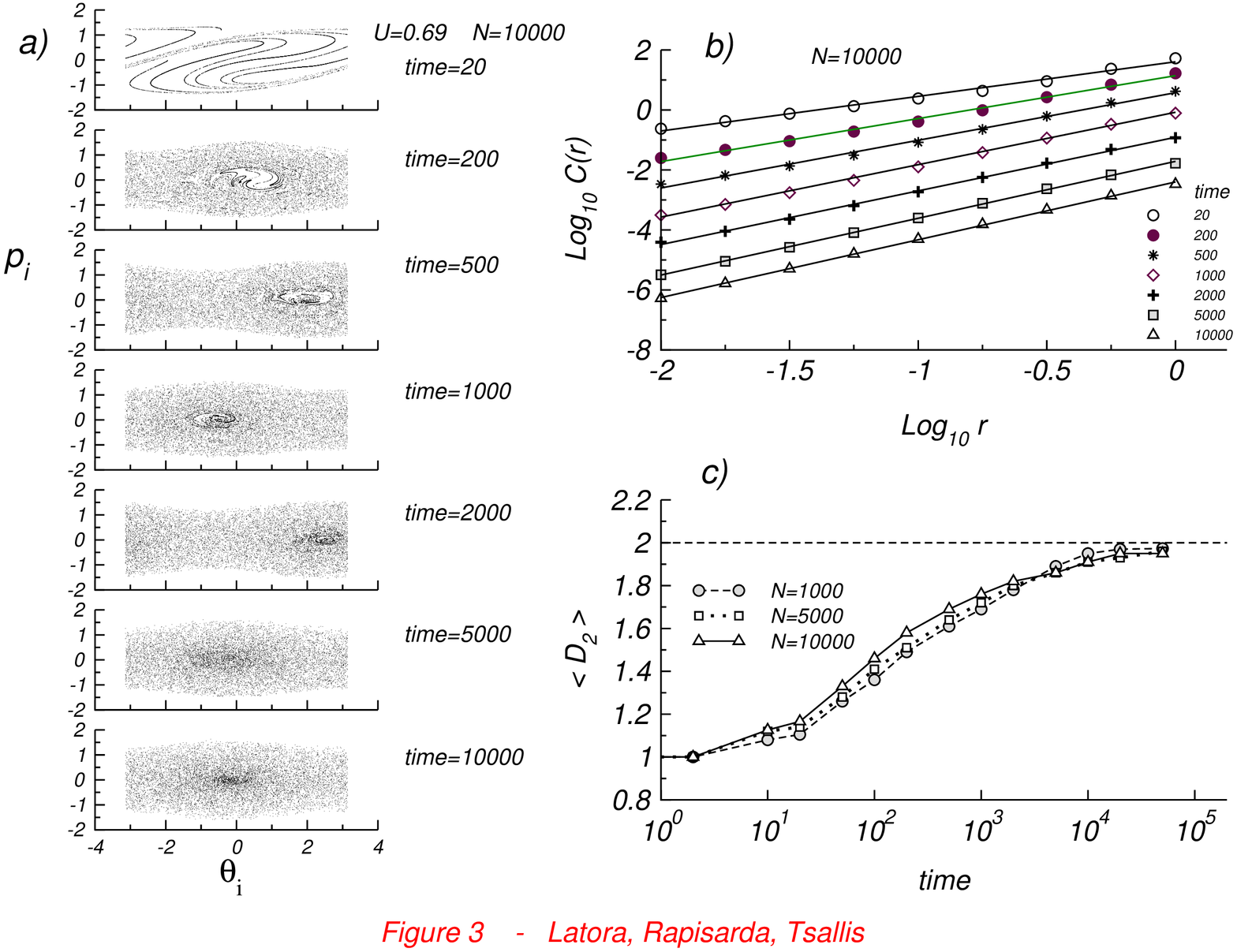,width=12truecm,angle=-90}
\end{center}
\caption{ Correlation in $\mu$-space. (a)  We show  the  $\mu$-space, i.e. 
the angle and momenta of the $N$ particles, for  $ U=0.69$   and  $ N=10000$  
 at  different  timescales, starting from WBIC. Although the initial 
configuration is uniform,  structures emerge and persist for a
 very long time before dissolving again at equilibrium. A way to
 measure these correlations is by means of the correlation 
integral\protect\cite{gras}
$ C(r)={1 \over N^2} \sum_{i,j}^N \Theta( r - d_{i,j} )$
 where $d_{i,j}$ is the Euclidean distance between 
two points of the $\mu$-space.  In general  $C(r)=r^{D_2}$, where $D_2$ is the
 correlation fractal dimension.   (b) By reporting  the logarithm of 
$ C(r)$ vs the logarithm of  $r$,  a  linear behavior over several decades is
 found.  The fractal dimension  thus extracted is reported in (c)  vs time
 (an average over 50 events is considered).  In the same time scale where
 we find the QSS, the correlation dimension is in between 1 and 2. The
 particles are fully spread in the  $\mu$-space only at equilibrium. 
As time increases, $D_2$  grows continuously from 1 to 2. 
}
\end{figure}

\end{document}